\documentclass[conference]{IEEEtran}
\IEEEoverridecommandlockouts
\usepackage{cite}
\usepackage{amsmath,amssymb,amsfonts}
\usepackage{algorithmic}
\usepackage{graphicx}
\usepackage{textcomp}
\usepackage{xcolor}

\ifCLASSOPTIONcompsoc
    \usepackage[caption=false, font=normalsize, labelfont=sf, textfont=sf]{subfig}
\else
\usepackage[caption=false, font=footnotesize]{subfig}
\fi

\def\BibTeX{{\rm B\kern-.05em{\sc i\kern-.025em b}\kern-.08em
    T\kern-.1667em\lower.7ex\hbox{E}\kern-.125emX}}
\begin{document}

\title{Achieving Robust Generalization for Wireless Channel Estimation Neural Networks by Designed Training Data}

\author{\IEEEauthorblockN{Dianxin Luan,~\IEEEmembership{Student Member,~IEEE,} John Thompson,~\IEEEmembership{Fellow,~IEEE}}\\
\IEEEauthorblockA{\textit{Institute for Digital Communications, School of Engineering, University of Edinburgh, Edinburgh, EH9 3JL, UK}\\
Email address : Dianxin.Luan@ed.ac.uk, john.thompson@ed.ac.uk}
}

\maketitle

\begin{abstract}
In this paper, we propose a method to design the training data that can support robust generalization of trained neural networks to unseen channels. The proposed design that improves the generalization is described and analysed. It avoids the requirement of online training for previously unseen channels, as this is a memory and processing intensive solution, especially for battery powered mobile terminals. To prove the validity of the proposed method, we use the channels modelled by different standards and fading modelling for simulation. We also use an attention-based structure and a convolutional neural network to evaluate the generalization results achieved. Simulation results show that the trained neural networks maintain almost identical performance on the unseen channels. 
\end{abstract}


\section{Introduction}
For future communication systems, artificial intelligence (AI) can help to improve the channel estimation task in wireless communications. ChannelNet \cite{soltani2019deep} and the untrained estimator proposed in \cite{balevi2020massive} are proved to achieve superior performance over conventional methods, which attracts a lot of current research interest. However, data-driven algorithms often degrade significantly on unseen channels. This prohibits the real-world implementation of neural networks because of insufficient stability on the practical channels. Online training can compensate for this degradation, but it requires additional consumption on both the latency and memory. It is inevitable to challenge the low-latency property and the low-complexity implementation. Moreover, the online trained neural networks have difficulties in predicting the channel gains at the data symbols precisely because the noise-free channel matrix of the complete packet is unavailable for training. This motivates us to investigate how to achieve robust generalization of general neural networks for channel estimation. However, the recent research only improves the generalization slightly for a narrow range of neural network applications \cite{zhang2017defense} \cite{wang2020high}. 

In this paper, we propose a method to design the training data for wireless channel estimation neural networks. To the authors' best knowledge, this is the first work that designs the training data for wireless channel estimation neural networks to achieve robust generalization. The trained neural networks can achieve robust generalization to the applicable channels constrained by the proposed method. It enhances the reliability of AI-assist channel estimation implementation. Therefore, by using the complete channel matrix as the training label, the trained neural networks perform precise interpolation in the time and frequency domain, which is the key advantage over the conventional methods. We deploy a low-complexity neural network (InterpolateNet \cite{luan2021low}) with 9,442 parameters and an attention-involved solution (HA02 \cite{9860803}) to test. The offline-trained InterpolateNet and HA02 are also simulated on the 3GPP TR 38.901 channels with varied delay spread to show the robustness with realistic channels. To support reproducibility, the MATLAB code can be downloaded at https://github.com/dianixn/ICC\_2023. 

Section \ref{System Architecture} presents the orthogonal frequency-division multiplexing (OFDM) baseband and frame structure based on the 5G New Radio (NR) standard along with the propagation channels. Section \ref{Conventional channel estimation methods and neural networks for simulation} introduces conventional channel estimation methods and describes the InterpolateNet and HA02 neural networks. Section \ref{Principle for designing the training data} proposes the design method for data generation. Section \ref{Simulation results} presents key simulation results. Section \ref{Conclusion} summarizes the key findings of this paper. 
\section{System Architecture}
\label{System Architecture}
\subsection{Baseband architecture}
\label{Baseband Architecture}
\begin{figure*}[htbp]
\centerline{\includegraphics[width=0.82\textwidth]{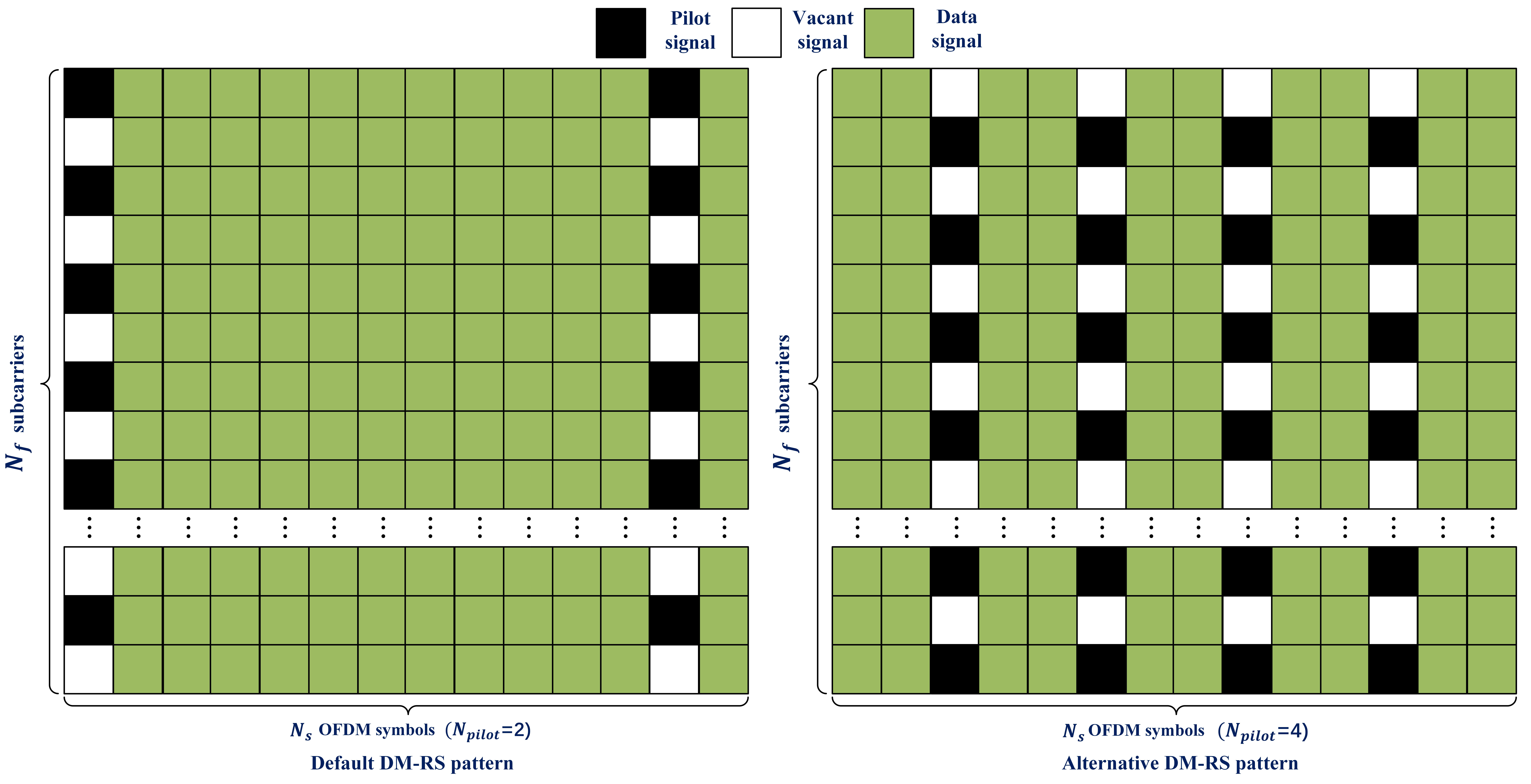}}
\caption{DM-RS pattern}
\label{DM-RS pattern}
\end{figure*}

This paper considers downlink wireless channel estimation for an OFDM cellular system. The source signal is processed by a Quadrature Phase Shift Keying (QPSK) modulator. The QPSK-modulated signals are assigned to the data subcarriers shown in Fig.~\ref{DM-RS pattern}. The default and alternative demodulation reference signal (DM-RS) patterns are defined in 3GPP TS 38.211 with $N_f$ subcarriers, $N_s$ OFDM symbols and $N_{pilot}$ pilot symbols. The pilot subcarriers and the vacant subcarriers of the pilot symbols are set to known values and 0 respectively. 
The inverse fast Fourier transform converts the OFDM symbols to the time domain signal. Then the cyclic prefix (CP, duration of $T_{CP}$) is added to the front of each symbol. Each slot is assigned by a new channel realization, which is assumed to be a multipath fading channel. Each path delay $\tau_m$ of channel satisfies that $\forall \tau_m \in [0, T_{CP}]$. At the single antenna receiver, the CP is removed and the received data is converted into the frequency domain using the fast Fourier transform operation. The received signal at the $k^{th}$ subcarrier and $l^{th}$ OFDM symbol, $Y(k, l)$, is 
\begin{equation}
Y(k, l) = H(k, l)X(k, l) + W(k, l), 
\end{equation}
where $H(k, l)$, $X(k, l)$ and $W(k, l)$ are the channel matrix, frequency domain OFDM signal and the additive white Gaussian noise at the $k^{th}$ subcarrier and the $l^{th}$ OFDM symbol. The received pilot signal is extracted to provide a channel reference for the complete packet. 
\subsection{Channel model}
\label{Channel model}
We consider single-input-single-output downlink scenarios operating at 2.1GHz with frequency spacing of 15kHz, and the duration of CP is $L_{CP}$ samples plus an implementation delay of 7 samples. The extended pedestrian A model (EPA), extended vehicular A model (EVA) and extended typical urban model (ETU) channels defined in 3GPP TS 36.101 and the customized channels with power delay profiles (PDP) provided in Table.~\ref{customized channels} are modelled by the generalized method of exact Doppler spread method \cite{patzold2009two} with maximum Doppler frequency 97 Hz for simulation. 
\begin{table}[htbp]
\caption{Customized channels}
\begin{center}
\begin{tabular}{|c|c|}
\hline
\multicolumn{2}{|c|}{Flat fading channel} \\
\hline
Path delay& 0 ns \\
\hline
Average path gain & 0.0 dB \\
\hline
\multicolumn{2}{|c|}{Defined channel 1 (DC1)} \\
\hline
Path delay& [0 50 100 200 400] ns \\
\hline
Average path gain & [0.0 -2.0 -4.0 -8.0 -16.0] dB \\
\hline
\multicolumn{2}{|c|}{Defined channel 2 (DC2)} \\
\hline
Path delay& [0 30 200 300 500 1500 2500 5000] ns \\
\hline
Average path gain & [-7.0 0 0 -1.0 -2.0 -1.0 -1.0 -5.5] dB \\
\hline
\multicolumn{2}{|c|}{Defined channel 3 (DC3)} \\
\hline
Path delay& [0 50 120 200 230 500 1600 2300 5000 7000] ns \\
\hline
Average path gain & [0.0 -1.0 -1.0 -1.0 -1.0 -1.5 -1.5 -1.5 -3.0 -5.0] dB \\
\hline
\multicolumn{2}{|c|}{Two-path fading channel} \\
\hline
Path delay& [50 5000] ns \\
\hline
Average path gain & [-3.0 -3.0] dB \\
\hline
\end{tabular}
\label{customized channels}
\end{center}
\end{table}
Clustered delay line (CDL) channels are modelled as link-level fading scenarios representing realistic channels, which have variable root mean square delay spreads ($\mathrm{DS}$). CDL-A, CDL-B and CDL-C are also implemented, which are constructed as non-line-of-sight channels following the 3GPP TR 38.901. In that standard, the delays can be scaled by 
\begin{equation}
\tau_{\mathrm{n, scaled}} = \tau_{\mathrm{n, model}}\mathrm{DS_{desired}}, 
\end{equation}
where $\tau_{\mathrm{n, scaled}}$ is the scaled delay value of $n^{th}$ cluster, $\tau_{\mathrm{n, model}}$ is the corresponding normalized model delay and $\mathrm{DS_{desired}}$ is the wanted $\mathrm{DS}$. For the carrier frequency of 2.1GHz, the range from 20ns (Short-delay profile, Indoor office) to 1148ns (Long-delay profile, Urban Macro) covers all the $\mathrm{DS_{desired}}$ values in 3GPP TS 38.901 Table 7.7.3-2. 
\section{Conventional channel estimation methods and neural networks for simulation}
\label{Conventional channel estimation methods and neural networks for simulation}
\subsection{Least squares (LS) method}
By minimizing the mean squared error (MSE) between $Y$ and $H \circ X$ at the pilot positions where $\circ$ denotes the Hadamard product, the frequency domain LS estimation is given by 
\begin{equation}
    \hat{H}_{LS} = \frac{Y_{Pilot}}{X_{Pilot}}, 
\end{equation}
Where $Y_{Pilot}, X_{Pilot} \in \mathbb{C}^{\frac{N_f}{2}\times N_{pilot}}$ denotes the received and transmitted pilot signals respectively and the mathematical division operation is performed element-wise. The estimation $\hat{H}_{LS}$ is then interpolated bilinearly to estimate the complete channel matrix $\in \mathbb{C}^{N_f \times N_s}$. 
\subsection{Minimum mean squared error (MMSE) method}
To minimize the distance between $H$ and $H_{LS}$, i.e. $\mathop{\arg\!\min_{H}}\Arrowvert H - \hat{H}_{LS}\Arrowvert^{2}_{2}$ at the pilot positions, the frequency domain MMSE estimate $\hat{H}_{MMSE}$ $ \in \mathbb{C}^{N_f \times N_{pilot}}$ is given by 
\begin{equation}
\label{MMSE}
    \hat{H}_{MMSE}(u) = R_{HH_{p}}(u)\left(R_{H_{p}H_{p}}(u) + \frac{\sigma_N^2}{\sigma_X^2}I\right)^{-1}\hat{H}_{LS}(u), 
\end{equation}
where $u$ denotes the index of the pilot OFDM symbol. $H(u) \in \mathbb{C}^{N_f}$ is the channel gain vector for the $u^{th}$ pilot OFDM symbol and $H_p(u) \in \mathbb{C}^{\frac{N_f}{2}}$ denotes the channel gain vector for the corresponding pilot symbol and subcarriers. The ratio (${\sigma_N^2}/{\sigma_X^2}$) is the numerical reciprocal of the signal-to-noise ratio (SNR) and $I$ is the identity matrix. The matrices $R_{HH_p}(u) = E\{H(u)H_{p}(u)^{H}\}$ and $R_{H_pH_p}(u) = E\{H_{p}(u)H_{p}(u)^{H}\}$ are the corresponding correlation matrices. The bilinear method is implemented to achieve interpolation for the channel gains at the data positions. 
\subsection{Neural networks for simulation}
InterpolateNet \cite{luan2021low} is a low-complexity neural network (9,442 parameters) that deploys bilinear interpolation to resize the in-processing features. HA02 \cite{9860803} is an encoder-decoder architecture which uses the transformer encoder \cite{vaswani2017attention} as the encoder and a residual convolutional architecture as the decoder respectively. HA02 exploits the self-attention mechanism to pre-process the LS estimate and focus on the critical features. Moreover, the outputs of both InterpolateNet and HA02 are $\hat{H} \in \mathbb{R}^{{N_f} \times N_{s} \times 2}$, which predict the channel matrix of the whole slot to achieve time and frequency interpolation. 
\section{Designing the training data for robust generalization}
\label{Principle for designing the training data}
The principle of neural networks is difficult to explain mathematically. The paper \cite{chan2022redunet} tries to explain the principle of convolutional neural networks but general interpretability of neural networks is still lacking. As we study the generalization for wireless channel estimation, we investigate the impacts of some intrinsic characteristics of the training data on the generalization properties of neural networks. 

It is proved that the choice of a sufficient rank and channel correlation \cite{edfors1998ofdm} affects the robustness to the variations in the channel statistics for MMSE filters, so that the MMSE channel estimator can generalize to these mismatched channels. The eigenvalue decomposition of the averaged channel auto-correlation $R_{HH}$ is given by 
\begin{equation}
R_{HH} = E\{HH^{H}\} = U\Lambda U^{H}, 
\label{equation}
\end{equation}
where $U$ contains the eigenvectors and $\Lambda$ is a diagonal matrix with the eigenvalues $\lambda$, which contains the power of the orthogonal channels \cite{edfors1998ofdm}. The first several elements of $\lambda$ are expected to contain the main power, which can ensure only a small performance loss when the designed estimator is simulated with some mismatched channel statistics. Moreover, a design for the worst correlation is declared to be robust to mismatch for a fixed finite impulse response estimator \cite{cavers1991analysis}. This hints to us that the generalization of the trained neural networks can be improved to resist channel mismatch by designing a channel to generate a proper training dataset. For that designed channel, the number and the magnitude of the corresponding main eigenvalues are expected to be larger than those of testing channels. This provides a high-rank and high-eigenvalue scenario (also suggested by \cite{edfors1998ofdm} \cite{cavers1991analysis}). 

Therefore, this paper investigates the impact of the path gains and the corresponding delays to create that desired channel to generate the training samples. As the phase of each path follows a uniform distribution and is uncorrelated to the corresponding power, we only design the PDP of the channel for training. We focus on the first $L_{CP}$ elements of eigenvalues because a path with a delay exceeding the CP duration should have an insignificant path gain. It also indicates that the maximum rank considered in this paper is $L_{CP}$. Moreover, the proposed method is applicable for general neural networks because no specific neural architecture is required. 

\begin{figure}[htbp]
\centerline{\includegraphics[width=0.45\textwidth]{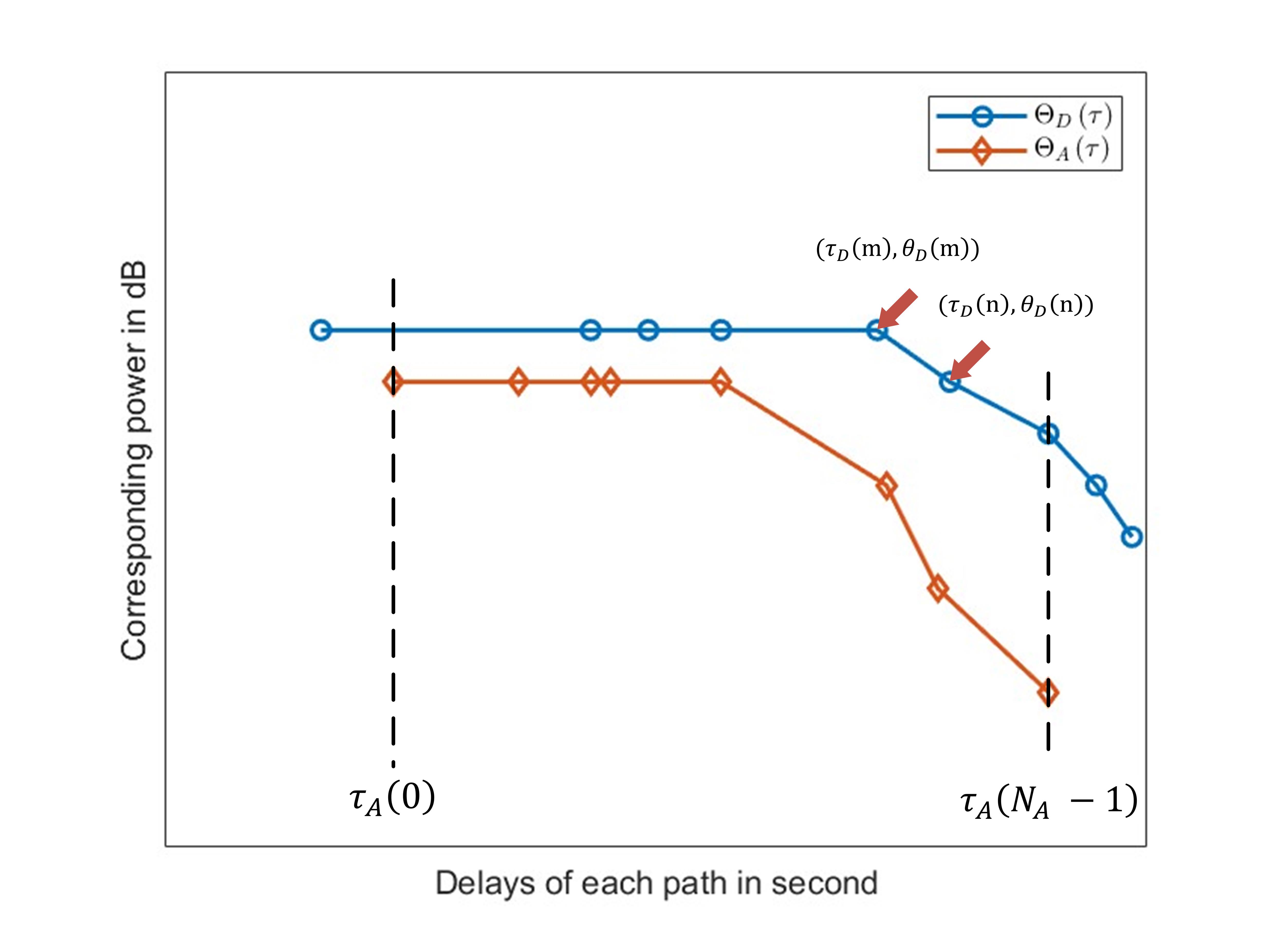}}
\caption{The designed PDP of $h_{D}$ (shown in blue) for the applicable channel $h_{A}$ (shown in red)(shown in red)}
\label{hT}
\end{figure}

To design the channel generating the training dataset, we assume that the designed channel $h_{D}$ is a Rayleigh fading channel modelled by \cite{patzold2009two} which has a fixed PDP that starts from 0dB approximately at delay 0ns and generally decays as the delay increases. This designed PDP is $\sum_{i=0}^{N_{D} - 1} \theta_{D}\left(i\right)\delta\left[n - \tau_{D}\left(i\right)\right]$, where $\tau_{D}(i)$ is the delay of the $i$th path, $\theta_{D}\left(\tau_{D}\left(i\right)\right)$ is the corresponding path gain and $N_{D}$ denotes the number of independent paths. The $\theta_{A}$, $\tau_{A}$ and $N_{A}$ represent the corresponding path gain and delay of an applicable channel $h_{A}$ (neural networks trained by $h_{T}$ can generalize to this channel). The corresponding PDP is $\sum_{j=0}^{N_{A} - 1} \theta_{A}\left(j\right)\delta\left[n - \tau_{A}\left(j\right)\right]$. The continuous function $\Theta_{D}\left(\tau\right)$ of the $h_{D}$ is obtained by linear interpolation between each two adjacent PDP's power-delay pairs. For arbitrary $\tau$ between the adjacent power-delay pairs $\left(\tau_{D}\left(m\right), \theta_{D}\left(m\right)\right)$ and $\left(\tau_{D}\left(n\right), \theta_{D}\left(n\right)\right)$ shown in Fig.~\ref{hT}, the value of $\Theta_{D}\left(\tau\right)$ between these two points is given by
\begin{equation}
    \frac{\Theta_{D}\left(\tau\right) - \theta_{D}\left(n\right)}{\tau - \tau_{D}\left(n\right)} = \frac{\theta_{D}\left(m\right) - \theta_{D}\left(n\right)}{\tau_{D}\left(m\right) - \tau_{D}\left(n\right)}. 
\end{equation}
$\Theta_{A}\left(\tau\right)$ is calculated in the same way as $\Theta_{D}\left(\tau\right)$. The applicable channel needs to satisfy the conditions 
\begin{gather}
    \forall \tau \in \left[\tau_{A}\left(0\right), \tau_{A}\left(N_{A} - 1\right)\right], \ \Theta_{A}\left(\tau\right) \leq \Theta_{D}\left(\tau\right). \\ 
    \tau_{A}\left(N_{A} - 1\right) \leq \tau_{D}\left(N_{D} - 1\right), \ \ N_{A} \leq N_{D}. 
\end{gather}
This should provide a worse correlation scenario and $h_{D}$ with more independent paths would have a larger rank than $h_{A}$. The SNR range of the designed channel is from 5dB to 25dB and the maximum Doppler shift is from 0Hz to 97Hz. The training parameters are given in Table.~\ref{Offline training parameters} for both the InterpolateNet and HA02. 
\begin{table}[htbp]
\caption{Training parameters}
\begin{center}
\begin{tabular}{|c|c|}
\hline
\textbf{Optimizer}& Adam\\
\hline
\textbf{Maximum epoch}& 100\\
\hline
\textbf{Initial learning rate}& 2e-3\\
\hline
\textbf{Drop period for learning rate}& every 20\\
\hline
\textbf{Drop factor for learning rate}& 5e-1\\
\hline
\textbf{Minibatch size}& 128\\
\hline
\textbf{L2 regularization}& 1e-7 \\
\hline
\end{tabular}
\label{Offline training parameters}
\end{center}
\end{table}
As regularization prevents overfitting as well as to improve generalization slightly, L2 regularization is set to be a very small value to lessen that positive effect. 
\section{Simulation results}
\label{Simulation results}
Equ.~(\ref{MSE}) defines the performance metric of MSE as 
\begin{equation}
    \mathrm{MSE}(\hat{H}, H) = \frac{1}{N_f N_s}\sum_{k=1}^{N_f} \sum_{l=1}^{N_s} {\left\Arrowvert\hat{H}_{kl} - H_{kl}\right\Arrowvert_{2}^{2}}, 
\label{MSE}
\end{equation}
where $H_{kl}$ is the real channel at $k$th subcarrier and OFDM $l$th symbol, and $\hat{H}_{kl}$ is the corresponding estimate. This paper sets $N_{f} = 72$, $N_{s} = 14$, a pilot value of (1+i) and $L_{CP} = 9$ for simulation. \textbf{Except for Section.~\ref{Control experimental}, both InterpolateNet and HA02 are only trained on a designed channel $h_{d}$} which has path delays of [0, 30, 200, 300, 500, 1500, 2500, 5000, 7000, 9000] ns and average path gains of [0, 0, 0, 0, 0, 0, -1, -1, -2, -4] dB. \textbf{This channel follows the proposed method, which ensures the trained InterpolateNet and HA02 generalize to the EPA, EVA, ETU, customized channels and CDL channels with normal delay spreads}. The training dataset comprises 125,000 samples, 95\% for training and 5\% for validation. The loss function for the InterpolateNet is MSE loss and for the HA02 is the Huber loss defined in equ.~(\ref{huber}). To average out Monte Carlo effects, each sample of the simulation curves is tested with 5,000 independent channel realizations. 
\begin{equation}
 L(a) =
\begin{cases}
\frac{1}{2}a^2& \text{if $|a| \leq 1$}\\
|a| - \frac{1}{2}& \text{otherwise}
\end{cases} 
\label{huber}
\end{equation}
\subsection{Generalization of the trained neural networks}
\label{Generalization of the trained neural networks to different Rayleigh channels}
To show the achieved generalization to the applicable channels, the trained neural networks are simulated on the channels modelled by \cite{patzold2009two} with an extended SNR range from 0dB to 30dB and the maximum Doppler shift from 0Hz to 97Hz. 
\begin{figure}[htbp]
\centerline{\includegraphics[width=0.5\textwidth]{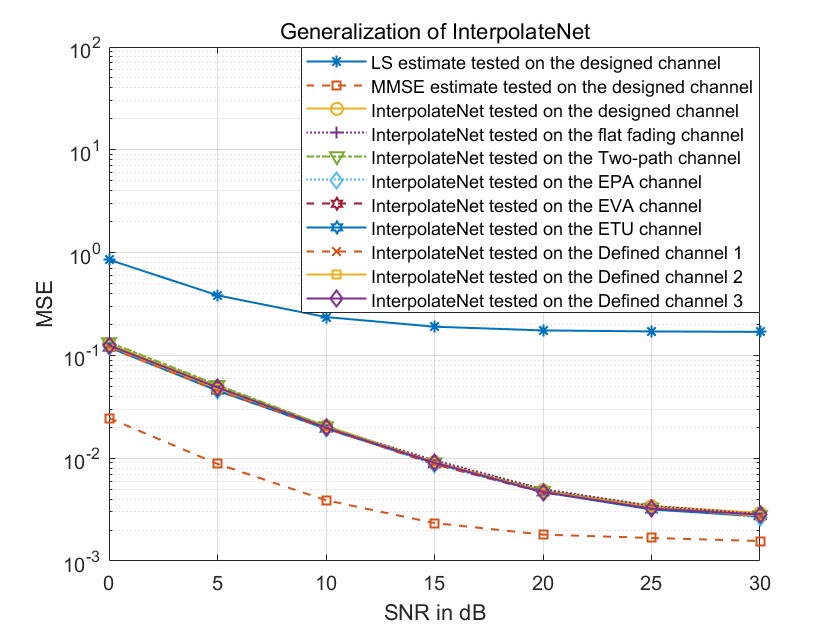}}
\caption{Generalization of the trained InterpolateNet}
\label{InterpolateNet Rayleigh}
\end{figure}
\begin{figure}[htbp]
\centerline{\includegraphics[width=0.5\textwidth]{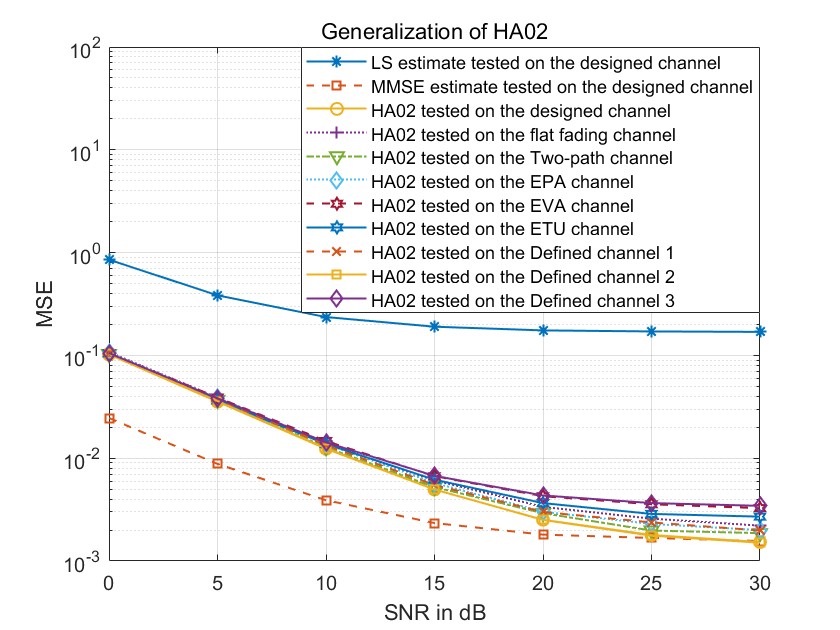}}
\caption{Generalization of the trained HA02}
\label{HA02 Rayleigh}
\end{figure}

Fig.~\ref{InterpolateNet Rayleigh} and Fig.~\ref{HA02 Rayleigh} compare the MSE results of the trained InterpolateNet and HA02 respectively on the applicable channels with the extended SNR. From Fig.~\ref{InterpolateNet Rayleigh}, \textbf{the trained InterpolateNet maintains an almost same performance over these channels with the MSE range from 0.0028 to 0.0030 at 30dB SNR}. \textbf{For the trained HA02, it also achieves robust generalization for SNR $\leq$ 15dB in Fig.~\ref{HA02 Rayleigh}. }However, the trained HA02 has slightly degraded generalization in the high SNR range because the attention-based encoder pre-processes the LS estimate to focus on the critical elements for the neural network decoder. The mismatch in the intrinsic properties of the channel introduces degradation, especially for the low-noise range as the channel noise realizes regularization. Therefore, the performance difference appears with the decreasing noise power where the MSE range is from 0.0009 to 0.0037 at 30dB SNR. Moreover, HA02 can outperform the MMSE method at SNR range from 20dB to 30dB on the DC2 channel because the capability of time interpolation is retained. 
\subsection{Validation of the proposed method}
\label{Control experimental}
\begin{figure}[htbp]
\centerline{\includegraphics[width=0.5\textwidth]{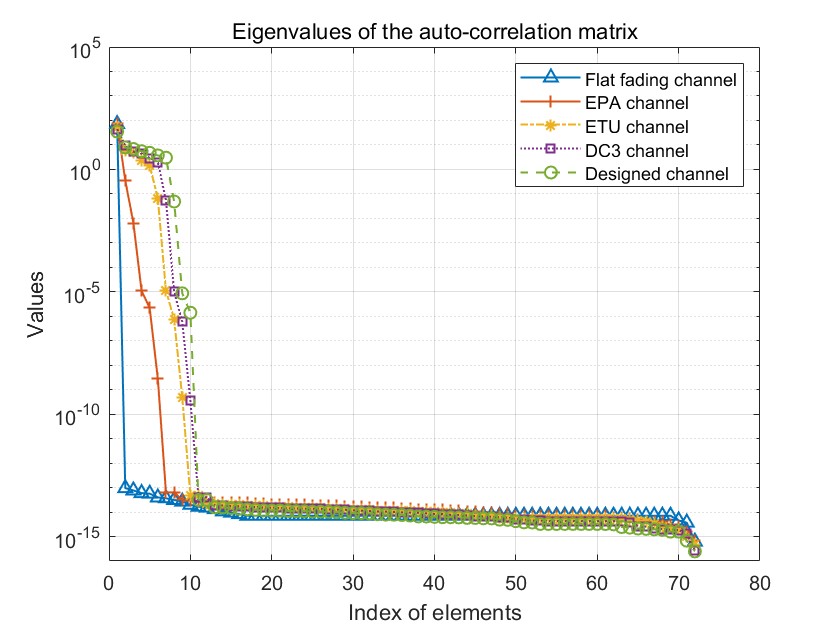}}
\caption{Eigenvalues of each channel's auto-correlation matrix}
\label{Singular}
\end{figure}

To justify the proposed method, we train the InterpolateNet and HA02 on different channels. The channel conditions are same as the designed channel. Fig.~\ref{Singular} shows the eigenvalues of the auto-correlation matrix for different channels. From what is proposed in Section.~\ref{Principle for designing the training data}, the neural networks trained on the channel with larger eigenvalues for the first $L_{CP}$=8 elements should generalize well to these with lower eigenvalues values, so for example the Designed channel will generalise well to all of the other channels that are shown. 
\begin{figure}[htbp]
\centerline{\includegraphics[width=0.5\textwidth]{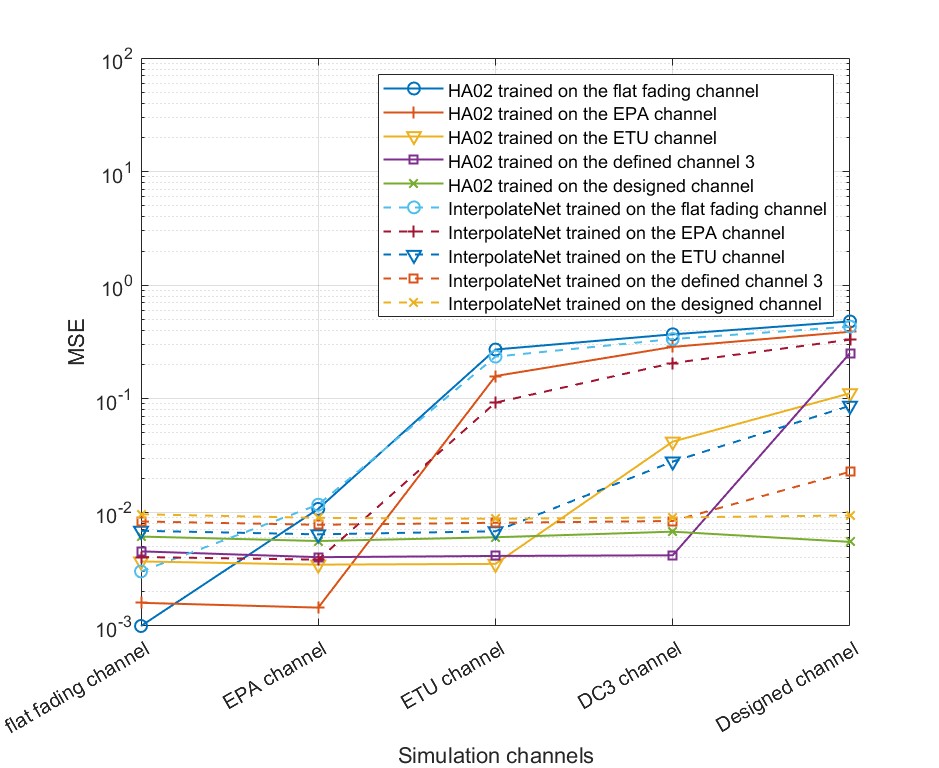}}
\caption{MSE results for InterpolateNet and HA02 trained on different channels}
\label{HA02 and InterpolateNet ablation}
\end{figure}

Fig.~\ref{HA02 and InterpolateNet ablation} compares the MSE results of the trained InterpolateNets and HA02s when tested on these channels with a fixed SNR of 15dB and the maximum Doppler shift identical with Section.~\ref{Generalization of the trained neural networks to different Rayleigh channels}. The MSE values for InterpolateNet and HA02 trained on the designed channel are from 0.0088 to 0.0096 and from 0.0056 to 0.0068 respectively, which proves robust generalization achieved over these channels. For the InterpolateNet and HA02 trained on the DC3 channel, the MSE value increase from 0.0084 to 0.0228 for InterpolateNet and from 0.0042 to 0.2524 for the HA02 respectively when the test channel switches from the DC3 channel to the designed channel. Otherwise, the trained InterpolateNet and HA02 have almost identical MSE performance over the flat fading, EPA and ETU channels. It is also observed that both the InterpolateNet and HA02 trained on the ETU channel only generalize to the EPA and flat fading channels well. EPA trained InterpolateNets and HA02 can only generalize to the flat fading channel and the flat fading channel provides limited generalization. These results match the hypothesis proposed. 
\subsection{Generalization to realistic channels}
\label{Generalization of the trained neural networks to the realistic channels}
\begin{figure}[htbp]
\centerline{\includegraphics[width=0.5\textwidth]{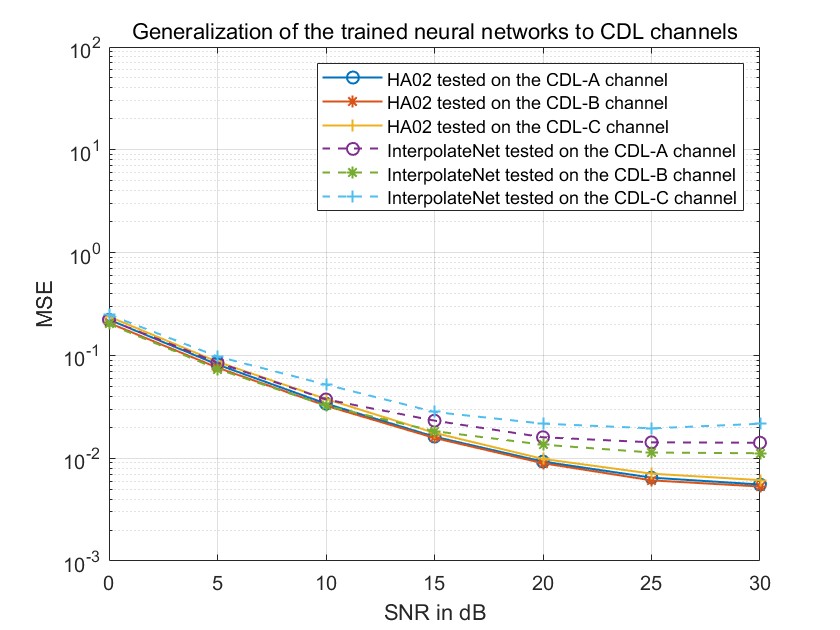}}
\caption{MSE results tested on the CDL channels with $\mathrm{DS_{desired}}$ of 30ns}
\label{Generalization to CDL channels (Default DM-RS pattern)}
\end{figure}

\begin{figure}[htbp]
\centerline{\includegraphics[width=0.5\textwidth]{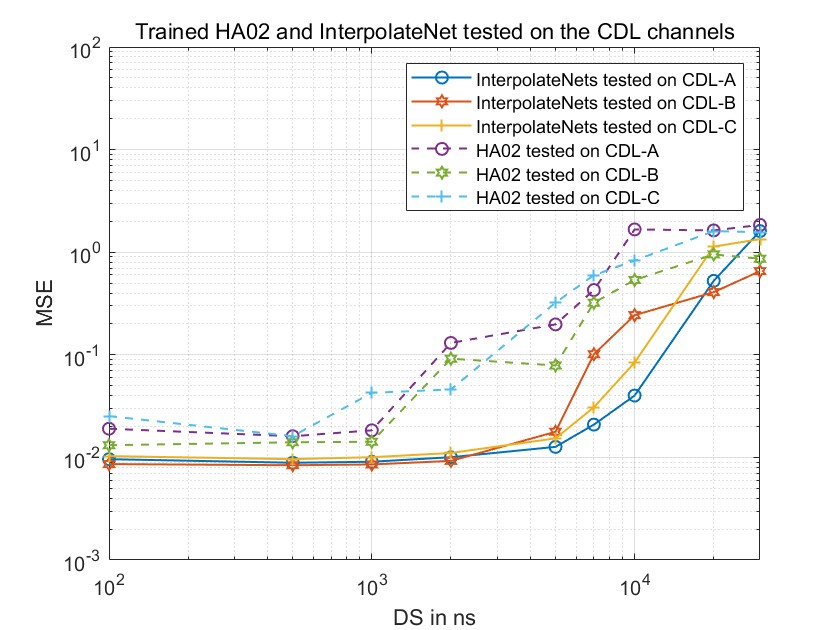}}
\caption{MSE results tested on the CDL channels with varying $\mathrm{DS_{desired}}$}
\label{1}
\end{figure}

To prove the generalization achieved to the realistic channels, the trained InterpolateNet and HA02 are simulated on the CDL-A, CDL-B and CDL-C channels by setting $\mathrm{DS_{desired}}$ to be 30ns for Fig.~\ref{Generalization to CDL channels (Default DM-RS pattern)} and the channel conditions are identical with Section.~\ref{Generalization of the trained neural networks to different Rayleigh channels}. To simulate the realistic scenarios by varying $\mathrm{DS_{desired}}$ to change the corresponding PDP, the trained InterpolateNet and HA02 are tested on these CDL channels with a fixed SNR of 20dB and the maximum Doppler shift identical with Section.~\ref{Generalization of the trained neural networks to different Rayleigh channels} for Fig.~\ref{1}. 

Fig.~\ref{Generalization to CDL channels (Default DM-RS pattern)} provides the MSE results of the trained InterpolateNet and HA02. Compared with Fig.~\ref{InterpolateNet Rayleigh} and Fig.~\ref{HA02 Rayleigh}, the performance of both trained neural networks has slight degradation when overcoming the mismatch of channel modelling between the designed channel and the CDL channels. Moreover, the degradation of the trained HA02 is relatively evident on the default DM-RS pattern. However, the trained InterpolateNet is still robust when tested on the CDL-A, CDL-B and CDL-C channels. At an SNR of 30dB, the variation of the trained InterpolateNet is negligible and the MSE range is from 0.0057 to 0.0066 while the variation of the trained HA02 is more severe (MSE from 0.011 to 0.021). Because of pre-processing the LS estimate to utilize the input sparsity, HA02 is more sensitive to the changes in simulation scenarios than InterpolateNet. 

Fig.~\ref{1} provides the MSE results when varying the $\mathrm{DS_{desired}}$ from 100ns to 30,000ns. For the trained InterpolateNet, the MSE values are almost same (from 0.0085 to 0.0151) when the $\mathrm{DS_{desired}}$ is from 100ns to 5,000ns. These MSE curves increase rapidly with the $\mathrm{DS_{desired}}$ above 9,000ns for all the CDL channels, which is reasonable because the corresponding delay spread significantly exceeds the duration of the CP. Moreover, the trained HA02 only maintains similar performance for the $\mathrm{DS_{desired}}$ from 100ns to 1,000ns, and degrades significantly for the $\mathrm{DS_{desired}}$ above 1,100ns. \textbf{Therefore, the trained InterpolateNet achieves robust generalization to these 3GPP TS38.901 $\mathrm{DS_{desired}}$ values and the trained HA02 maintains a reliable performance over the DS range. }
%
%
%
\subsection{Adaptation to alternative DM-RS pattern}
\label{Adaptation to the alternative DM-RS pattern}
To show the adaptation to different DM-RS patterns, the InterpolateNet and HA02 are trained on the $h_{d}$ because of the pilot positions of the alternative DM-RS pattern are changed, and then simulated with the same settings as Fig.~\ref{1} to investigate the effect of varying $\mathrm{DS_{desired}}$. 
%
%
%
\begin{figure}[htbp]
\centerline{\includegraphics[width=0.5\textwidth]{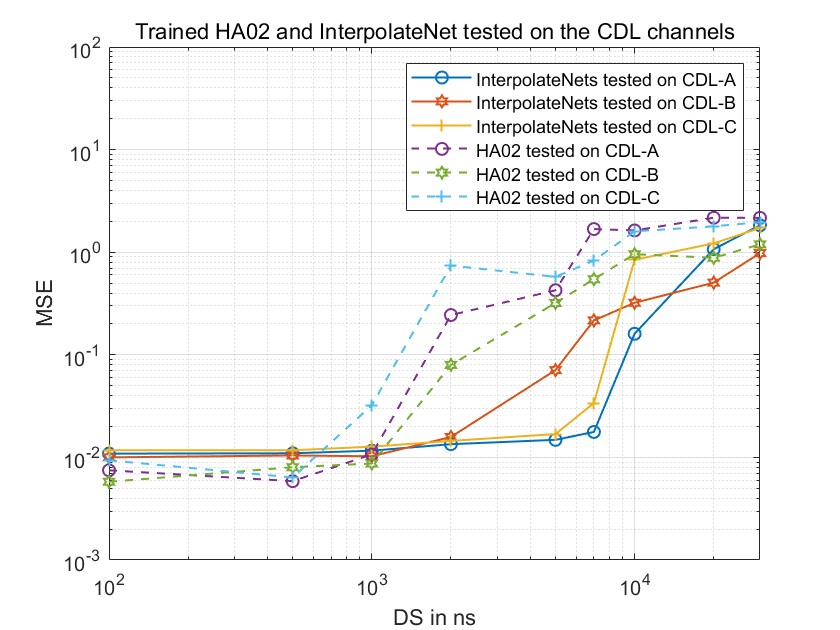}}
\caption{MSE results tested on the CDL channels with the varying $\mathrm{DS_{desired}}$ (Alternative DM-RS pattern)}
\label{2}
\end{figure}
%

From Fig.~\ref{2}, the trained InterpolateNet still has an identical performance with MSE values from 0.0100 to 0.0163 for the $\mathrm{DS_{desired}}$ from 100ns to 1,200ns, and degrades with the increasing of $\mathrm{DS_{desired}}$. Moreover, the trained HA02 slightly outperforms the trained InterpolateNet for the low $\mathrm{DS_{desired}}$ from 100ns to 700ns. Therefore, the impacts caused by substituting different DM-RS patterns are insignificant. 
\section{Conclusion}
\label{Conclusion}
We propose a method to design neural network training samples, to achieve robust generalization for wireless channel estimation. It is based on a worst case power delay profile design and ensures the trained neural networks maintain an almost identical performance on the applicable channels. From the simulation results, the trained InterpolateNet and HA02 generalize to the test channels robustly. The trained InterpolateNet and HA02 are also shown to generalize to realistic channels with only a small compromise in MSE. This method avoids the requirement that neural network solutions need online training procedures to adjust the parameters. By following the proposed method to train the neural networks, the trained neural networks can still predict the complete channel matrix precisely in realistic environments. 
\ifCLASSOPTIONcaptionsoff
  \newpage
\fi

\bibliographystyle{IEEEtran}

\bibliography{Reference}

\end{document}